\PassOptionsToPackage{pdftex,
pdfencoding=auto,
pdfnewwindow=true,
pdfusetitle=true,
bookmarks=true,
bookmarksnumbered=true,
bookmarksopen=true,
pdfpagemode=UseThumbs,
bookmarksopenlevel=1,
pdfpagelabels=true,
breaklinks=true,
colorlinks=true,
}{hyperref}
\documentclass[amsmath,amssymb,superscriptaddress,twocolumn,notitlepage,nofootinbib]{revtex4-2}

\usepackage[utf8]{inputenx}
\usepackage[english]{babel}
\usepackage{graphicx}
\usepackage[usenames,dvipsnames]{xcolor}
\definecolor{purple}{RGB}{128,0,128}
\definecolor{ultramarine}{RGB}{63, 0, 255}
\definecolor{medblue}{RGB}{0, 0, 100}
\definecolor{googleblue}{RGB}{34, 0, 204}
\definecolor{panblue}{RGB}{0,24,150}
\definecolor{carmine}{RGB}{150, 0, 24}
\definecolor{gray}{RGB}{150, 150, 150}
\definecolor{darkgreen}{RGB}{0, 80, 0}
\usepackage{amsmath}
\usepackage{amssymb}
\usepackage{url}
\usepackage{hyperref}
\usepackage[]{hyperref}
\hypersetup{linkcolor=carmine,citecolor=darkgreen,urlcolor=googleblue,anchorcolor=OliveGreen}

\usepackage{microtype}
\microtypecontext{spacing=nonfrench}
\microtypesetup{
expansion={true,nocompatibility},
protrusion={true,nocompatibility},
activate={true,nocompatibility},
tracking=true,
kerning=true,
spacing={true}
}
\usepackage[all=normal,floats=tight,mathspacing=tight,wordspacing=tight,paragraphs=normal,tracking=tight,charwidths=tight]{savetrees}
\everypar=\expandafter{\the\everypar\loosness=-1 }
\usepackage[style=american,autopunct=true]{csquotes} 
\usepackage{mathtools}

\usepackage{soul}

\newcommand{\ket}[1]{\left|#1\right\rangle}
\newcommand{\bra}[1]{\langle#1|}

\newcommand\blfootnote[1]{%
  \begingroup
  \renewcommand\thefootnote{}\footnote{#1}%
  \addtocounter{footnote}{-1}%
  \endgroup
}

\newcommand{\bracket}[1]{\langle#1\rangle}

\usepackage{bbold}
\usepackage[flushleft, neveradjust]{paralist}

\usepackage{subfigure}

\usepackage{verbatim}
\usepackage{verbatimbox}

\graphicspath{{./images/}}

\begin{document}
\onecolumngrid
\twocolumngrid

\title{ Network Nonlocality with Separable Measurements}

\author{Emanuele Polino}
\email{e.polino@griffith.edu.au , emanuele.polino@gmail.com}
\affiliation{Queensland Quantum and Advanced Technologies Research Institute, Centre for Quantum Computation and Communication Technology, Griffith  University, Yuggera Country,   Brisbane,  Queensland,  4111  Australia}

\author{Davide Poderini}
\affiliation{Università degli Studi di Pavia, Dipartimento di Fisica, QUIT Group, via Bassi 6, 27100 Pavia, Italy}
\affiliation{International Institute of Physics, Federal University of Rio Grande do Norte, 59078-970, Natal, Brazil}

\author{Giorgio Minati}

\author{Giovanni Rodari}

\affiliation{Dipartimento di Fisica - Sapienza Universit\`{a} di Roma, P.le Aldo Moro 5, I-00185 Roma, Italy}

\author{Rafael Chaves}
\email{rafael.chaves@ufrn.br}
\affiliation{International Institute of Physics, Federal University of Rio Grande do Norte, 59078-970, P. O. Box 1613, Natal, Brazil}
\affiliation{School of Science and Technology, Federal University of Rio Grande do Norte, Natal, Brazil}

\author{Fabio Sciarrino} 
\email{fabio.sciarrino@uniroma1.it}

\affiliation{Dipartimento di Fisica - Sapienza Universit\`{a} di Roma, P.le Aldo Moro 5, I-00185 Roma, Italy}

\begin{abstract}

Quantum correlations in networks with independent sources have revealed novel forms of nonclassical behavior. While entanglement in the sources is a necessary ingredient, the role played by entanglement in the measurements remains largely unexplored.  In particular, all existing demonstrations of full network nonlocality, certifying the nonclassicality of every source in the network, have relied on entangled measurements performed at a central node with no inputs. In this work, we construct an explicit strategy that does not rely on entangled measurements, yet still achieves full network nonlocality. Our approach is based on separable measurements augmented with bidirectional classical feedforward. We further show that this same class of measurements can give rise to another recently proposed form of network nonlocality, the minimal network nonclassicality, which ensures that the observed correlations cannot be attributed to any fixed subset of nonclassical sources within the network. Finally, building on a recently developed certification framework, we quantify the amount of device-independent randomness that can be extracted from full network nonlocal correlations under different measurement strategies. Beyond their foundational significance, our results also offer a practically attractive route toward experimental implementations of network nonlocality, as they remove the need for entangled measurements.
\end{abstract}

\blfootnote{Corresponding authors:}
\maketitle

\section{Introduction}

Quantum nonlocality represents one of the most distinctive departures of quantum theory from classical physics~\cite{bell1964einstein,Brunner}, and can be understood as the violation of classical causal constraints~\cite{wood2015lesson,chaves2015unifying,wiseman2017causarum,wolfe2020quantifying}. Beyond its foundational interest, nonlocality is a key resource for device-independent (DI) quantum information processing~\cite{pironio2016focus,poderini2022ab}, where conclusions can be drawn without assumptions on the internal functioning of the devices, playing a central role for secure quantum networks~\cite{wehner2018quantum}.

Motivated by both fundamental and practical considerations, increasing attention has been devoted to quantum networks that extend beyond the standard bipartite Bell scenario, involving multiple parties connected by shared quantum resources~\cite{gisin2023quantum}. In this setting, multipartite nonlocality has become an active area of research, with several recent experimental demonstrations~\cite{mao2024certifying,villegas2024nonlocality,broad2024,pepper2024scalable,webb2024experimental,weinbrenner2024certifying,hu2025observation}. Of particular interest are networks with multiple independent sources, which give rise to nontrivial causal structures and impose strong constraints on classical correlations~\cite{tavakoli2022bell}. These scenarios have been extensively investigated theoretically and realized experimentally, especially in photonic platforms~\cite{saunders2017experimental,carvacho2017experimental,sun2019experimental,poderini2020experimental,li2022testing,wu2022experimental,carvacho2022quantum,agresti2021experimental,suprano2022experimental,d2023machine,polino2023experimental,wang2024experimental,wang2023certification,wang2025simultaneous}, and have enabled applications including communication protocols~\cite{zukowski1993event,azuma2023quantum}, randomness certification~\cite{minati2026randomness,alanon2025certifying}, and self-testing~\cite{renou2018self,agresti2021experimental,vsupic2023quantum,sekatski2023partial,sarkar2025self,sarkar2026universal}. They have also revealed new phenomena, such as nonclassical correlations without external inputs~\cite{branciard2012bilocal,fritz2012beyond,renou2019genuine,chaves2021causal,polino2023experimental,pozas2023proofs,baumer2025exploring,wang2024experimental,lauand2025minimal}, and evidence for the necessity of complex numbers in quantum theory~\cite{renou2021quantum,li2022testing,chen2022ruling,wu2022experimental}.

\begin{figure}[t!]
    \centering
    \includegraphics[width=\columnwidth]{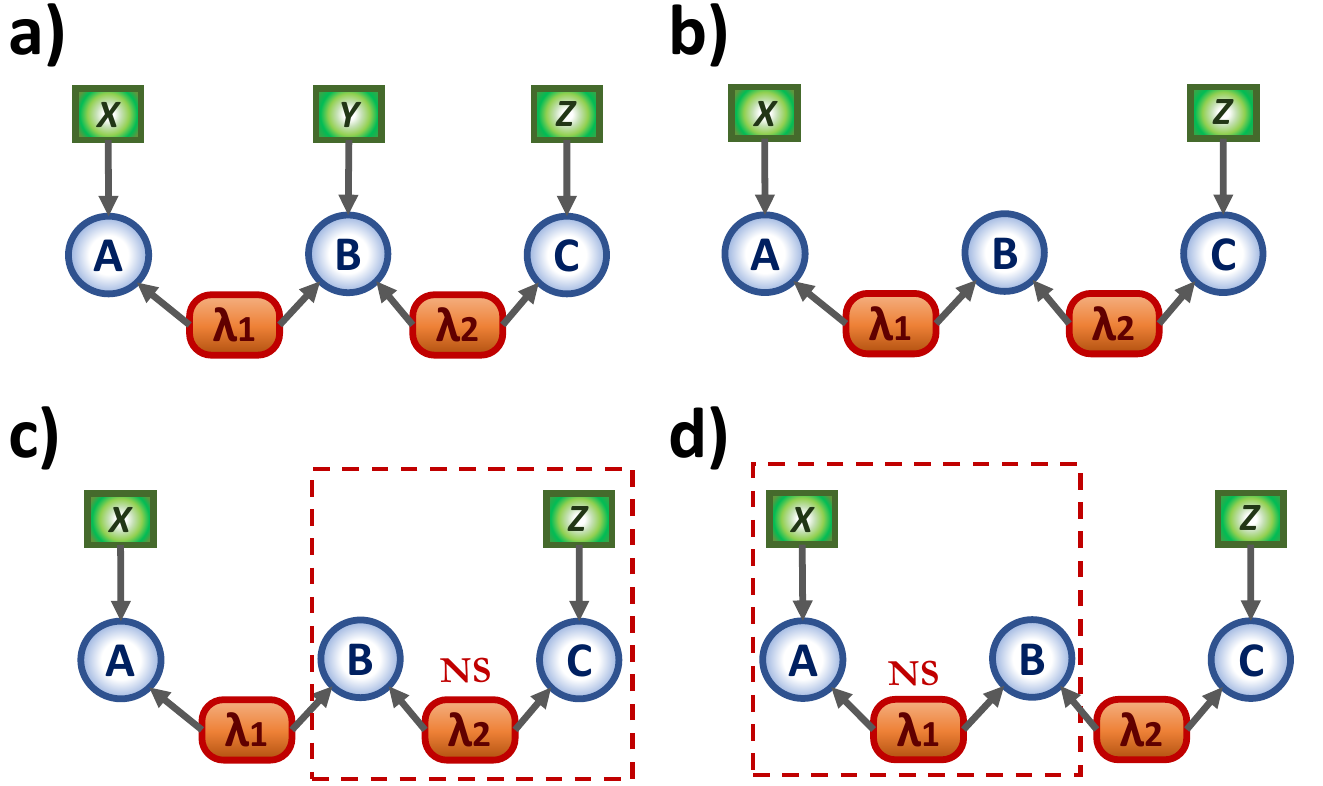}
    \caption{Bilocality scenarios. \textbf{a)} with an external input per measurement station; \textbf{b)} with external inputs only in the external nodes; \textbf{c)} when the second source $  \lambda_2$ shares a no-signaling (NS) resource (red dotted square); \textbf{d)} when the first source $  \lambda_1$ shares an NS resource.}
    \label{fig:bilodags}
\end{figure}

Since in network scenarios some measurement stations have to process subsystems coming from different independent sources, quantum strategies for demonstrating nonlocality in networks may involve measurements that are either entangled or separable. A measurement is entangled when its associated operator corresponds, up to normalization, to an entangled state. Entangled measurements~\cite{jozsa2003entanglement} play a central role in protocols such as teleportation and entanglement swapping~\cite{bennett1993teleporting,zukowski1993event,hu2023progress}, and have motivated extensive studies of genuinely quantum forms of network nonlocality~\cite{gisin2019entanglement,renou2019genuine,tavakoli2021bilocal,vsupic2022genuine,huang2022entanglement,pozas2022full,wang2023certification,gu2023experimental,wang2024experimental,del2024iso,pauwels2025classification,mao2024certifying,wei2024nonprojective,ding2025symmetricquantumjointmeasurements,wang2025simultaneous,pauwels2025symmetric,pauwels2025multiqubit,ustun2025fusion}. However, entangled measurements are not always required to violate network locality. For instance, in the bilocal scenario, where two independent sources distribute subsystems to three parties (Fig.\ref{fig:bilodags}), violations of bilocal classical models can be achieved using separable measurements~\cite{branciard2010characterizing,branciard2012bilocal}. Similar observations apply to star-network configurations~\cite{tavakoli2014nonlocal,andreoli2017maximal,poderini2020experimental} and to the triangle network, where feedback-based strategies can reveal nonclassical correlations~\cite{fritz2012beyond,polino2023experimental}.

Despite these results, the general relation between the type of measurements performed at the nodes and the forms of nonlocality that can be demonstrated in complex networks remains unclear~\cite{tavakoli2022bell,cavalcanti2023fresh}. A central open question is therefore which forms of network nonlocality can be demonstrated without invoking entanglement within the measurement stations. Addressing this issue is relevant both for clarifying the origin of network nonclassicality and for practical implementations. Entangled measurements are challenging to realize efficiently in quantum optics~\cite{calsamiglia2002generalized,bianchi2025bell}, and avoiding them can improve robustness and practical feasibility, for instance by circumventing fourth-order interference and enabling hybrid photonic networks operating at different wavelengths~\cite{carvacho2022quantum}.

In this work, we take a step forward in addressing this question by identifying strategies that, using measurements that do not involve entanglement, can violate two forms of network nonlocality: full network nonlocality (FNN)~\cite{pozas2022full}, and Minimal Network Nonclassicality (MNN)~\cite{ciudad2024escaping}. FNN refers to correlations that certify, in a device-independent manner, that all sources in the network are nonclassical (see also Ref.\cite{luo2024hierarchical} for a hierarchy of network correlations). These correlations have been experimentally realized in a variety of networks~\cite{haakansson2022experimental,wang2023certification,huang2022entanglement,gu2023experimental,wang2025simultaneous} and theoretically generalized~\cite{munshi2025device}. The ability to certify the nonclassicality of all sources is an important feature of full network nonlocality, with direct applications to the certification of resources for quantum information protocols, such as quantum key distribution \cite{mukherjee2026full}.

While previous demonstrations of FNN in the bilocal scenario without inputs at the central node (Fig.~\ref{fig:bilodags}-c,d) relied on entangled measurements, we show that this requirement can be relaxed. We present a scheme based on separable measurements supplemented with classical feedback that suffices to demonstrate FNN in this setting.

Then we also show that this type of measurement is suitable for demonstrating MNN, that is, a recently introduced notion of network nonclassicality complementary to FNN~\cite{ciudad2024escaping}. MNN characterizes the set of nonclassical correlations that can be simultaneously explained by all bilocal scenarios in which only one of the two sources is nonclassical. This notion enforces a form of delocalized nonclassicality, where the observed correlations cannot be attributed to any fixed subset of nonclassical sources, thereby “escaping” any embedding into standard Bell scenarios~\cite{ciudad2024escaping}.

We also analyze the certifiable randomness generated by the separable measurement strategy and compare it with scenarios employing entangled measurements at the central node. 

\begin{figure*}[t!]
    \centering
    \includegraphics[width=\textwidth]{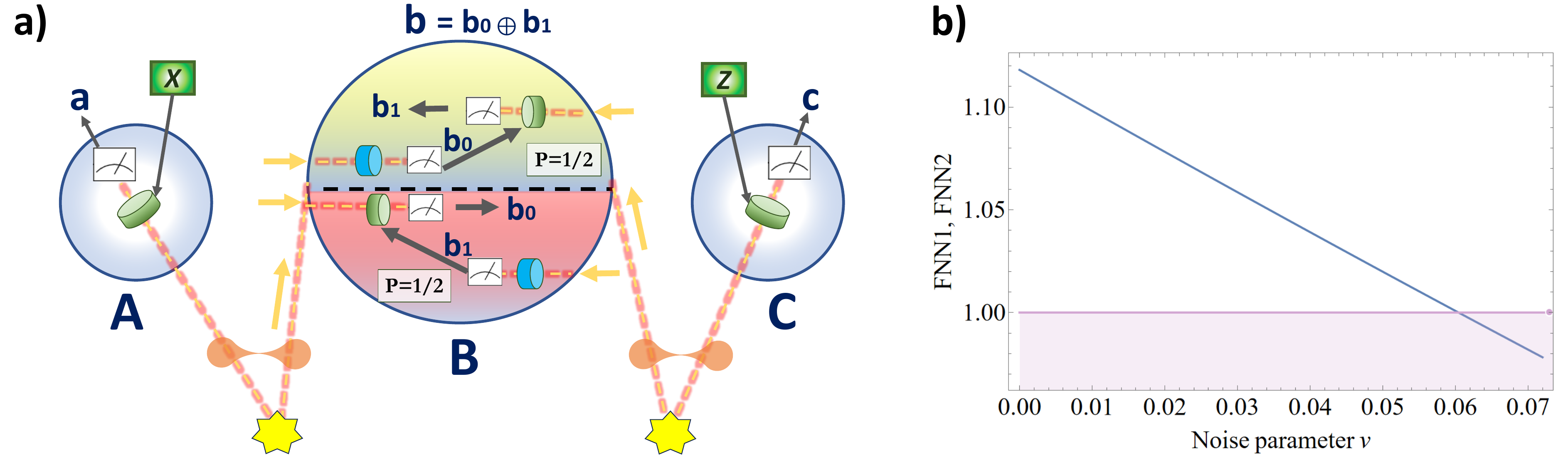}
    \caption{Measurement entanglement-free strategy. \textbf{a)} Scheme of the entanglement-free, feedback-based strategy to demonstrate FNN in the bilocal scenario. The measurement in the central node consists of two alternating feedback-based measurements, each occurring half of the time.  The half circle in yellow shows the feedback from the subsystem of the source connecting $A$ and $B$, determining the measurement on the subsystem from the source connecting $B$ and $C$. \textbf{b)} Values of ${FNN}_1={FNN}_2$ in Eq.\eqref{eq:full} as a function of the white noise parameter of the two identical Werner states shared among the nodes of the scenario. Purple shadow represents the region of the inequalities that can be reached by classical-no-signaling hybrid models (no FNN).}
    \label{fig:results}
\end{figure*}

\section{Full Network Nonlocality and Minimal Network Nonclassicality in the bilocal scenario}

The scenario we consider in this work is the bilocality network, which involves two independent sources connecting three measurement stations as represented in the causal structure of Fig. \ref{fig:bilodags}-b. This network has been realized in several photonic implementations showing the quantum violation of bilocal models using separable~\cite{andreoli2017experimental,poderini2020experimental} or entangled~\cite{carvacho2017experimental,saunders2017experimental,sun2019experimental} measurements in the central node.
Full network nonlocality has been introduced for the first time in this scenario~\cite{pozas2022full}. 
This stronger notion of network nonlocality is generated across the entire network and guarantees that no subset of sources composing the network can be described by local classical hidden variables, even if the other sources are constrained only by the no-signaling principle. 
More specifically, full network nonlocal correlations in the bilocal scenario cannot be explained by any models of the following form (Fig.\ref{fig:bilodags}-c):
\begin{equation}\label{fullmodel}
	p(a,b,c|x,z)=\int d\lambda\; p(\lambda) \;  p(a|x,\lambda) \; p^{NS}(b,c|\lambda,z)  \; ,
\end{equation}
where $p(\lambda)$ refers to a classical hidden variable describing one of the sources, $p^{NS}(b,c|\lambda,z)$ is the bipartite distribution of the variables constrained only by no-signaling, namely \mbox{$\sum_b p(b,c|\lambda,z)=p(c|z)$} and similarly inverting the roles of $c$ and $a$ (Fig.\ref{fig:bilodags}-d):
\begin{equation}
\label{fullmodel2}
 p(a,b,c|x,z)=\int d\lambda \;  \; p(\lambda) \; p(c|z,\lambda) \; p^{NS}(a,b|\lambda,x) \; .
\end{equation}

If only one of the two sources is classical and regardless of the possible post-quantum no-signaling nature of the other source, at least one of the following network Bell-like inequalities has to hold \cite{pozas2022full}:
\begin{equation}
 \label{eq:full}
 \begin{multlined}
 \begin{dcases}
\textit{FNN}_1= -\bracket{A_1B C_2}-\bracket{A_2B}+ \\
+\bracket{C_2}\, \left[ \bracket{A_1B}+\bracket{A_2BC_2}+\bracket{C_2} \right]\leq 1 \\
\\
\textit{FNN}_2= -\bracket{A_1BC_2}+\bracket{BC_1}+\\
+\bracket{A_1}\,\left[\bracket{BC_2}-\bracket{A_1BC_1}+\bracket{A_1}\right]\leq 1,
 \end{dcases}
 \end{multlined}
\end{equation}
where $\bracket{A_iBC_j}=\sum_{a,b,c}(-1)^{a+b+c}p(a,b,c|x=i,z=j)$, the variables $a,\; b,\; c \in \{0,1\}$ are dichotomic outcomes, and similarly for the other correlators and single-party mean values.
Note that the central measurement station $B$ only performs a single measurement without external input (see Fig.\ref{fig:bilodags}-b for the corresponding causal structure).
Quantum systems can simultaneously violate the inequalities \eqref{eq:full} using suitable entangled measurements \cite{pozas2022full}. In Refs.~\cite{gu2023experimental,wang2025simultaneous}, partial Bell state measurements~\cite{pan1998greenberger} were used to experimentally violate full network nonlocality, while Elegant Joint Measurements~\cite{gisin2017elegant,gisin2019entanglement} were employed in Ref.~\cite{huang2022entanglement}.
As discussed in the next section, FNN can also be demonstrated without entangled measurements.

In Ref.\cite{ciudad2024escaping}, another form of network nonclassicality, called Minimal Network Nonclassicality (MNN), has been introduced as a sufficient condition to identify correlations in a quantum network that are not traceable back to the standard Bell theorem.
More specifically, we say that a distribution is MNN if it cannot be explained by purely classical sources but is compatible with every causal model where a single arbitrary source is nonclassical. Practically, MNN correlations need to be compatible with both the scenarios in Figs.\ref{fig:bilodags}-c-d. By definition, the set of MNN correlations is disjoint from the FNN correlation set. 
Not relying on any specific subset of parties exploiting a nonclassical common cause, the nonclassicality arising from MNN has to be delocalized, escaping any embedding into the standard Bell scenario~\cite{ciudad2024escaping}. 
In the bilocal scenario, a distribution can be certified as MNN via a set of quadratically constrained quadratic programs \cite{lauand2023witnessing}, and an MNN distribution has been provided by Ref.\cite{ciudad2024escaping} using an entanglement-swapping distribution with entangled measurement at the central node, together with an additional necessary amount of noise. We show that the separable measurement strategy that we introduce in the next section is able to generate MNN distributions.

\section{Measurement-entanglement-free strategy for network nonlocality}
For the FNN case, the most straightforward way to certify that both sources in the bilocal scenario are nonclassical without resorting to entangled measurements would be to perform two independent Bell tests, one on each source. Such a strategy, however, necessarily requires at least one external input at the central node $B$, as illustrated in Fig.~\ref{fig:bilodags}-a. In contrast, full network nonlocality (FNN) has been defined for the scenarios depicted in Fig.~\ref{fig:bilodags}-b–d \cite{pozas2022full}, where node $B$ has no free inputs and the nonclassicality emerges from the network structure itself. This causal configuration, which demands less external randomness, is also more resource-efficient from the perspective of quantum information processing. In this setting, however, parallel Bell tests are not possible, and standard separable measurements at the central node do not lead to a violation of Eq.~\eqref{eq:full} (see Appendix A).

Here, to overcome this limitation, we introduce a different setup based on feedback-based measurements~\cite{wiseman2009quantum}, which are not entangled but only classically correlated. See Fig.\ref{fig:results}-a for the conceptual scheme of the strategy. To illustrate, consider the scenario with two maximally entangled two-qubit states pairwise shared among the three stations: 
$\ket{\Psi^{-}}_{ABC}= \ket{\Psi^{-}}_{AB_0} \otimes \ket{\Psi^{-}}_{B_1C}$, where $ |\Psi^{-}\rangle = (|01\rangle -|10\rangle)/\sqrt{2}$.
Each of the two external nodes receives one qubit from a different source and performs a standard projective measurement on it.
Conversely, the central node receives two qubits, one from each source, and performs a separable measurement on them. To achieve a demonstration of FNN, a feedback-based strategy alternatively uses the bits generated by the subsystem of the first source $\rho_{AB}$, which we indicate as $B_0$, to choose the measurement on the subsystem $B_1$  from  $\rho_{BC}$ and vice versa. The two directions of the feedback 
alternate half of the time. 
This kind of double feedback measurement can be formalized by the following measurement operator $M^B_b$:
\begin{eqnarray}
\label{eq:FritzPOVMs}
 && M^{B_0 B_1}_{(b_0,b_1)}= \frac{1}{2}M^{B_0}_{b_0}\otimes M^{B_1}_{b_1|b_0} + \frac{1}{2} M^{B_0}_{b_0|b_1}\otimes M^{B_1}_{b_1}\nonumber \\  \\ [-0.3em] \nonumber
 && M^B_b = \sum_{b_0\oplus b_1=b}  M^{B_0 B_1}_{(b_0,b_1)}
\end{eqnarray}

where $M^{B_0}_{b_0}\otimes M^{B_1}_{b_1|b_0}$ represents the measurements in the central node $B$ where the outcome $b_0$ from the subsystem $B_0$ influences the measurement $M^{B_1}_{b_1|b_0}$ performed on the subsystem $B_1$, while    $M^{B_0}_{b_0|b_1}\otimes M^{B_1}_{b_1}$ represents the opposite case where the outcome $b_1$ from  $B_1$  influences the measurement $M^{B_0}_{b_0|b_1}$ performed on $B_0$. The outcome of the central node $B$ is given by the sum modulo 2 of the bits relative to the two subsystems $b_0 \oplus b_1$.
This measurement strategy exploits the outcome obtained from one source to determine the measurement performed on the qubit originating from the second source. It is conceptually reminiscent of the adaptive measurements required to realize the Fritz distribution in the triangle scenario \cite{fritz2012beyond,fritz2016beyond,polino2023experimental,chaves2021causal}. Importantly, however, our approach does not rely on Fritz-type arguments rooted in the standard Bell scenario. The adaptive structure is employed solely to achieve a violation of the inequality in Eq.~\eqref{eq:full}, which itself is derived independently of any Fritz-like construction.

This strategy can be viewed as two intertwined Bell tests, where the outcome associated with one subsystem from a given source effectively acts as an input for the measurement performed on the subsystem from the other source. There is, however, a crucial difference from simply running two completely parallel Bell tests. In the latter case, at least one freely chosen input, independent of both sources, is required. In the scenario considered here, any variable that determines the measurement in the central node is causally connected to the sources and may, in principle, be fully determined by them. This also applies to a hypothetical hidden variable that would “select” the direction of the feedback at the central node. As a result, the protocol requires strictly less external seed randomness than two independent Bell tests.

Using this setup, a violation of Eq.\eqref{eq:full} can be achieved by reaching the simultaneous values $\textit{FNN}_1 = \textit{FNN}_2 \simeq 1.11803$, numerically optimized as a function of the ten single qubit projectors required to test FNN with the introduced strategy (four for the external nodes, and six for the central node). See Appendix A 
for the details on the projector measurements used to achieve the optimal value of FNN. Note that, using entangled measurements, it is possible to achieve higher values for the violations:  $\textit{FNN}_1 = \textit{FNN}_2 =\frac{1}{2}(1+\sqrt{2}) \simeq 1.20711 $ \cite{pozas2022full}.

We emphasize that the proposed measurement scheme, which removes the need for entangled measurements, significantly relaxes the experimental requirements for photonic implementations of full network nonlocality. In particular, it eliminates the need for picosecond-level synchronization and fourth-order interference, thereby opening the possibility of employing completely independent photon sources without relying on probabilistic entangling operations.

In Fig.\ref{fig:results}-b we show the range of noises in the shared identical isotropic Werner states, $ \rho_{AB}=\rho_{BC} = (1-\nu) \ket{\Psi^-}\bra{\Psi^-} + \nu \; \mathbb{1}/4$, with $ |\Psi^{-}\rangle = (|10\rangle -|01\rangle)/\sqrt{2}$ and $\mathbb{1}/4$ the identity matrix in 4-D representing a fully mixed state. The critical value of the white noise parameter $\nu \simeq 0.06043$ below which the states, measured by the proposed scheme, can show FNN (Fig.\ref{fig:results}-b).

In the following, we show how the feedback-based measurements can be used to generate MNN correlations in the bilocal scenario in Fig.\ref{fig:bilodags}-b. 
In order to do that, we need to introduce some noise in the measurement in the $B$ node, by using the following two-outcome (outcome $b=0,1$) POVM $\Pi^{B}_{b}$ generalizing Eq.~\eqref{eq:FritzPOVMs}:
\begin{eqnarray}
\label{eq:MNNFritz}
\nonumber & & \Pi^{B_0 B_1}_{(b_0, b_1)}= \\ \nonumber
 & &\frac{\alpha_0+\alpha_1}{2}  \Big( p\;M^{B_0}_{b_0}\otimes M^{B_1}_{b_1|b_0} + (1-p) M^{B_0}_{b_0|b_1}\otimes M^{B_1}_{b_1} \Big) \\  \nonumber 
&& + \frac{(1-\alpha_{b_0\oplus b_1})}{2} \mathbf{I}\;,\\  \\ [-0.6em] \nonumber
&&  \Pi^{B}_{b}= \sum_{b_0\oplus b_1=b} \Pi^{B_0 B_1}_{(b_0,b_1)}\;,
\end{eqnarray}
where $\mathbf{I}$ is the identity matrix in the 4-dimensional space of the two qubits.
Here, the wirings work exactly as before, the $p$ parameter controls the proportion of the mixture of feedback strategies (in the previous section, we considered $p=1/2$), and $(1-\alpha_{0,1})$ denotes the different noise coefficients for the two outcomes of the measurement at node $B$, and $0<\alpha_{0,1}<1$.
Using this measurement at the central node $B$ and two singlet states as sources, we tested the resulting distribution via the Quadratically Constrained Quadratic Problem (QCQP) described in Appendix~B.
By varying $\alpha_{0,1}$ appropriately, this strategy can produce distributions that exhaust all the possibilities regarding its reproducibility with (non)classical sources.

\begin{figure}[t!]
    \centering
    \includegraphics[width=\columnwidth]{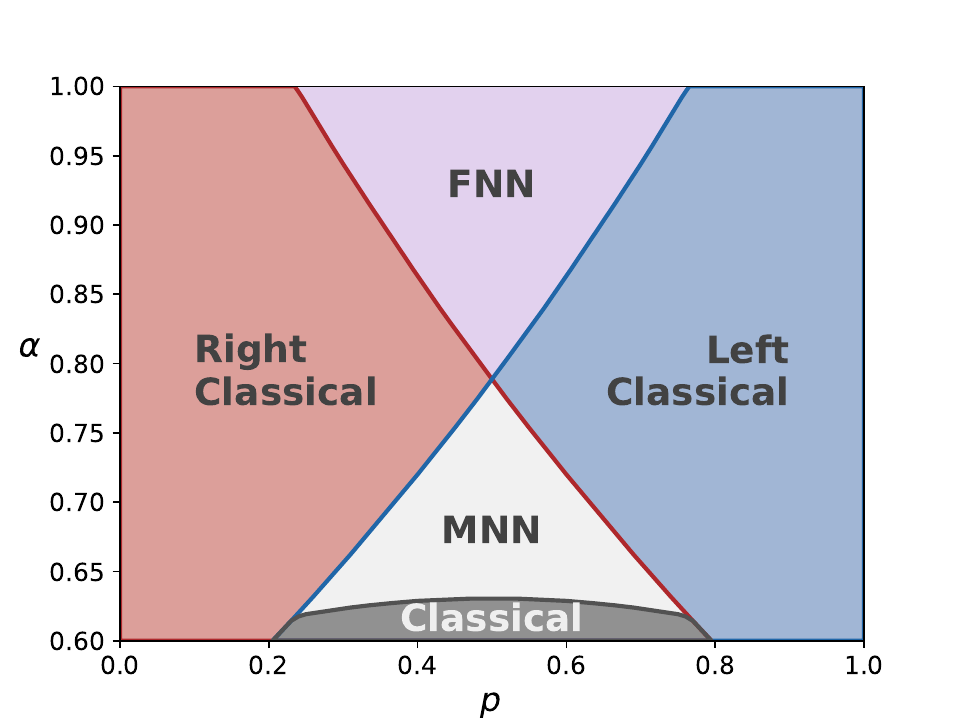}
    \caption{Nonclassicality with feedback-based measurements. Sets of nonclassicality as a function of the parameters $p$ and $\alpha=\alpha_1$ ($\alpha_0=1$) enabled by the measurement strategy in Eq.\eqref{eq:MNNFritz}, when two maximally entangled states are produced by the two sources of the bilocal scenario.}
    \label{fig:alpha_p_regions}
\end{figure}

As expected, moving $p$ away from $p=1/2$, for some level of noise $\alpha_{0,1}$, makes it necessary for either the source on the right or the left to be classical.
Instead, when $p=1/2$, the distribution can only be classical, MNN, or FNN.
In particular, we see that for $\alpha_0 = \alpha_1$, the distribution never belongs to the MNN, while this happens as soon as we introduce asymmetric noise (in the outcomes).
In particular, we can fix $\alpha_0=1$ and vary $\alpha_1$ to obtain a class of strategies that reproduces correlations that cover all the range of possibilities.
Fig.\ref{fig:alpha_p_regions} shows the results of the numerical study for this case and the different kinds of nonclassicality as a function of the mixture proportion $p$ and the noise $\alpha_1$. 

The regions of different colors refer to different kinds of correlations achievable by the measurement strategy in Eq.\eqref{eq:MNNFritz}. 
These include the FNN region (purple), obtained when $p$ is sufficiently balanced; the MNN region (white), arising when the noise ratio, $\alpha\equiv \alpha_0 / \alpha_1$, is sufficiently unbalanced; fully classical correlations (grey), appearing in the high-noise regime with balanced $p$; and the Right Classical (red) and Left Classical (blue) regions in the remaining parts of the parameters space.

\section{Entangled versus separable measurements: a randomness certification comparison}
\begin{figure}[ht!]
    \centering
    \includegraphics[width=0.45\textwidth]{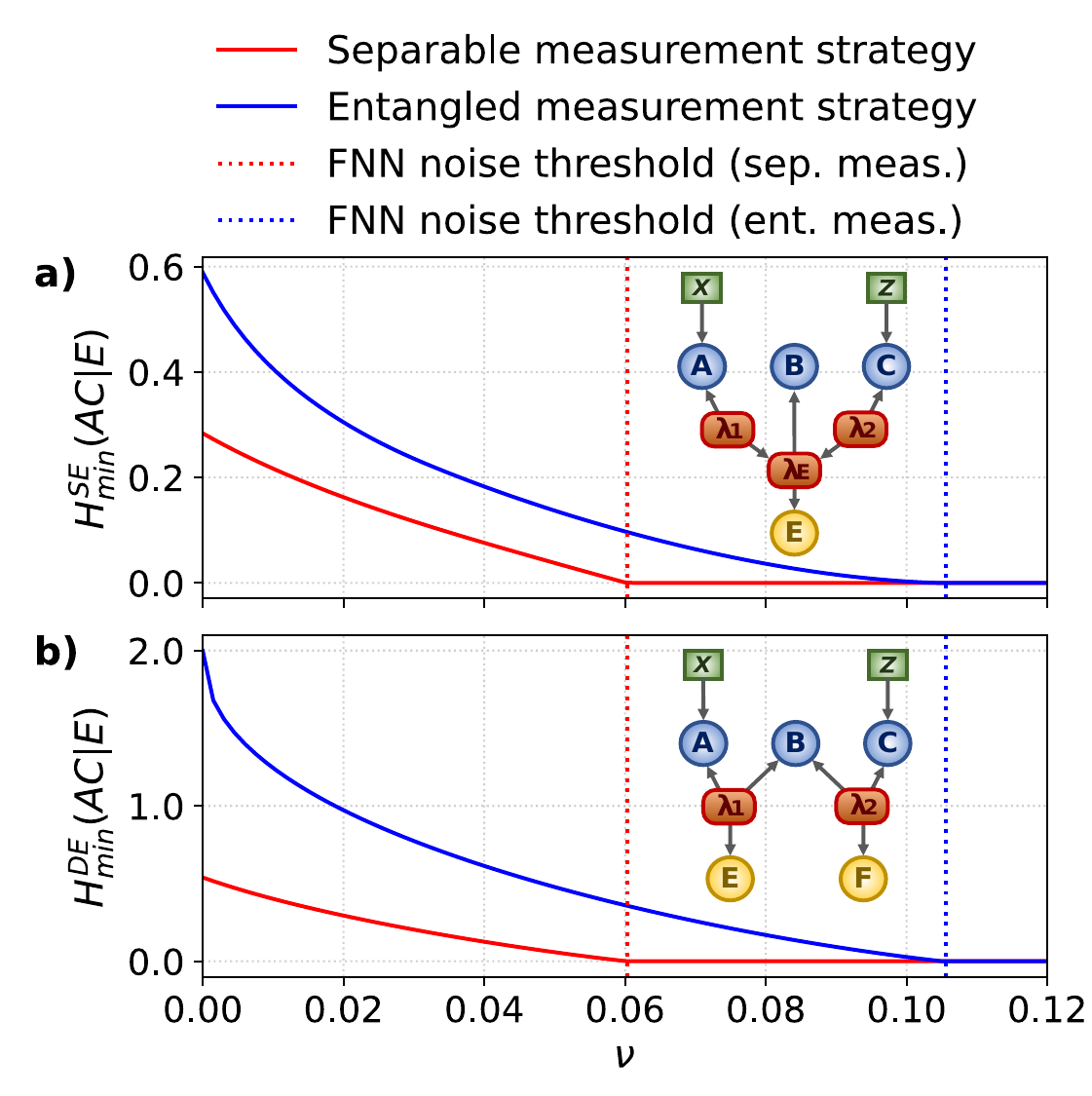}
    \caption{Randomness certification comparison between the entangled (blue curves) and entanglement-free (red curves) measurement schemes for FNN. Two-party min-entropy $H_{\mathrm{min}}^{\mathrm{SE/DE}}(AC|E)$ as a function of the states' white noise parameter $\nu$ in the \textbf{a)} SE and \textbf{b)} DE eavesdropper's configurations.}
    \label{fig:rand}
\end{figure}

A central and increasingly active research direction in quantum networks concerns the extension of randomness generation protocols to multipartite scenarios, where new challenges and opportunities emerge \cite{woodhead2018randomness,grasselli2023boosting,wooltorton2025expanding,li2024randomness,broad2024,minati2026randomness,alanon2025certifying}.
In this section, we compare the amount of randomness that can be certified from full network nonlocal correlations using our feedback-based separable measurement strategy with that obtainable via entangled measurements.
For this purpose, we employ the technique introduced in Ref.~\cite{minati2026randomness}. We consider two attack models, referred to as the Strong-Eavesdropper (SE) scenario and the Double-Eavesdropper (DE) scenario, whose causal structures are shown in the insets of Fig.~\ref{fig:rand} a-b. The former represents the worst-case scenario within the bilocal network, as the eavesdropper has access to both independent sources and, through an additional latent node ($\lambda_E$ in Fig.~\ref{fig:rand}-a), full knowledge of Bob's measurement outcomes, independently of the implemented strategy. In contrast, in the DE scenario, the eavesdropper has only separate access to the sources, represented by the two independent nodes $E$ and $F$ in the inset of Fig.~\ref{fig:rand}-b, and is therefore restricted to measurements of the form $E_e \otimes F_f$.

Given a specific scenario, the amount of randomness in the measurement outcomes can be quantified in a DI manner by optimizing over all possible eavesdropping strategies (see Appendix~C for details on the numerical methods). As a figure of merit, we consider the \textit{min-entropy} $H_{\mathrm{min}}(AC|E)$, which provides a lower bound on the number of certifiable random bits extractable from the measurement outcomes of the outer nodes $A$ and $C$ in the presence of an eavesdropper $E$. This quantity is particularly relevant for two reasons. First, the bilocal network serves as the prototypical structure in a quantum repeater scenario, where two distant parties rely on an intermediate node, making the outputs of the outer nodes the primary targets for an eavesdropper. Second, by focusing on the external nodes, we can assess the indirect influence that entangled measurements at the central node have on the outcome statistics of $A$ and $C$.

In Fig.\ref{fig:rand}, we report the numerical min-entropy obtained in both the SE and DE scenarios, comparing the results achieved when the central node performs either the feedback-based separable measurements described in Eq.\eqref{eq:FritzPOVMs} or the entangled Elegant Joint Measurements reported in \cite{pozas2022full}. As expected, a non-zero amount of randomness is certified only when the white noise level $\nu$ on the quantum states $\rho_{AB}$ and $\rho_{BC}$ is low enough to achieve full network nonlocality. Within the SE eavesdropper scenario, the separable-measurement strategy allows us to certify up to 0.288 bits of randomness, a value surpassed by the entangled measurement strategy, which achieves a maximum of $0.588$ random bits. The advantage of the entangled strategy is even more significant in the DE scenario, where the eavesdropper is limited to separable measurements. In this case, we can certify $0.539$ bits with the feedback-based separable strategy, whereas performing Elegant Joint Measurements allows us to reach the maximum certifiable amount of randomness, i.e., $2$ bits. This implies that, in this situation, the eavesdropper can only uniformly guess the measurement outcomes in $A$ and $C$. 

These results suggest that, in the context of randomness certification, performing entangled measurements in the central node is indeed a valuable resource, both in terms of noise tolerance and maximum amount of certified randomness. 

\section{Discussion}

Future quantum communication infrastructures are expected to rely on complex networks that distribute information securely through intrinsically quantum resources. Motivated by both foundational questions and technological applications, significant effort has been devoted to the realization of multipartite quantum networks in a variety of configurations~\cite{saunders2017experimental,carvacho2017experimental,sun2019experimental,poderini2020experimental,li2022testing,wu2022experimental,carvacho2022quantum,agresti2021experimental,suprano2022experimental,d2023machine,polino2023experimental,wang2024experimental,wang2023certification,ho2022entanglement,mao2022test,cao2022experimental,huang2022experimental,mao2024certifying,villegas2024nonlocality,broad2024,pepper2024scalable,webb2024experimental,pickston2023conference,proietti2021experimental}.
A central problem in this context is identifying the minimal physical resources required to generate nonclassical correlations in such networks. In particular, it is essential to clarify when entangled measurements are genuinely necessary. This question is relevant both for understanding the structure of quantum correlations and for guiding experimental implementations. Nonetheless, much of the existing literature on network nonlocality has focused on scenarios involving entangled measurements~\cite{tavakoli2022bell}, and experimental efforts toward their photonic realization are advancing steadily~\cite{hou2018deterministic,bayerbach2023bell,yan2023universal,d2025boosted,akin2025faithful,hauser2025boosted}. In contrast, comparatively less attention has been given to separable measurements supplemented by classical feedback.

In this work, we investigate the capabilities of such measurements that can be constructed from local separable measurements combined with classical communication between subsystems at the measurement station, without invoking entangled resources. We present an explicit strategy in which bidirectional feedback enables the generation of full network nonlocality (FNN) in the bilocal scenario without entangled measurements. This is consistent with the definition of FNN in this setting, which guarantees that all network sources must be nonclassical but does not certify the presence of entanglement in the central measurement node~\cite{pozas2022full}. If one aims to certify entanglement in the measurement station itself, additional or different tests would be required. For example, one could supplement the protocol with a standard bipartite Bell inequality between the external nodes in an entanglement-swapping configuration. Alternatively, the magnitude of the violation of Eqs.~\eqref{eq:full} may provide partial information about the presence of entanglement in the central measurement, a question that deserves further investigation.
We further show that the same feedback-based strategy, supplemented with a suitable measurement noise, enables the observation of MNN \cite{ciudad2024escaping}. In this way, our approach qualitatively spans a broad range of network nonclassicality correlations. Notably, the protocol achieves this without relying on entanglement swapping, in direct alignment with the original MNN paradigm, thereby demonstrating that distinct forms of network nonclassicality can be accessed through a single, resource-efficient architecture.

Finally, we compared entangled and feedback-based implementations from the perspective of device-independent randomness certification, using the framework introduced in Ref.~\cite{minati2026randomness}. Our analysis indicates that FNN generated via entangled measurements can achieve higher certifiable randomness rates compared with the considered adaptive separable strategies, highlighting a quantitative advantage of measurement entanglement in this context.

More broadly, our results motivate further exploration of generalized feedback-based measurement strategies and their role in characterizing different notions of network nonlocality. In particular, it remains an open question whether correlations produced by such protocols should be regarded as genuinely network nonlocal under various definitions, or whether they can ultimately be reduced to standard bipartite Bell nonlocality in disguise (see, e.g., Refs.~\cite{ciudad2024escaping,vsupic2022genuine}). Clarifying this point may contribute to a more refined understanding of nonclassical correlations in complex quantum networks.

\section*{Acknowledgments}
The authors thank A. Pozas-Kerstjens for useful discussions and comments.
The authors acknowledge support from FARE Ricerca in Italia QU-DICE Grant n. R20TRHTSPA, QU-BOSS (QUantum advantage via nonlinear BOSon Sampling, Grant Agreement No. 884676), the PNRR MUR Project No. PE0000023-NQSTI. This work was supported by the Australian Research Council; E.P. is a recipient of an Australian Research Council Discovery Early Career Researcher Award (DE250100762). R. C. acknowledges the Simons Foundation (Grant No. 1023171, R.C.), the Brazilian National Council for Scientific and Technological Development (CNPq, Grants No.403181/2024-0, and 301687/2025-0), the Financiadora de Estudos e Projetos (Grant No. 1699/24 IIF-FINEP)

\section*{Appendix A: Details on the measurement strategy}
\label{app:measu}

The strategy for generating FNNs involves a feed-forward approach, with feedback directions changing at different times; indeed, any strategy without this double-directional feedback would be explainable by one of the models in Eqs. \eqref{fullmodel} and \eqref{fullmodel2}  as we briefly show now. Consider a bilocal scenario in which one subsystem $B_0$ of the intermediate node $B$ (from the source connecting $A$ and $B$) is measured first, with outcome $b_0$ fed forward to the second measurement, producing $b_1$. The observed distribution can be written as: 
\begin{equation*}
p(a,b,c|x,z)=\sum_{b_0\oplus b_1=b} \mathrm{Tr}\!\left[\rho_1 \otimes \rho_2 \;. \; M^A_x \otimes M^{B_0}_{b_0}\otimes M^{B_1}_{b_1|b_0} \otimes M^C_z \right].
\end{equation*}
Because the measurement on $B_0$ is fixed, the expression factorizes as
\begin{equation*}
\begin{split}
&\mathrm{Tr}\!\left[\rho_1 \;. \; M^A_x \otimes M^{B_0}_{b_0} \right] \;
\mathrm{Tr}\!\left[\rho_2 \;. \; M^{B_1}_{b_1|b_0} \otimes M^C_z  \right]=\\
&= p(a,b_0|x)\; p(b_1,c|b_0,z)
\end{split}
\end{equation*}
yielding to
\begin{equation*}
p(a,b,c|x,z) = \sum_{b_0\oplus b_1=b} p(a,b_0|x)\; p(b_1,c|b_0,z)\; .
\end{equation*}
This decomposition always holds in the presence of the feed-forward $B_0 \to B_1$. Since any distribution $p(a,b_0|x)$ can be reproduced using a classical source (it would correspond to a Bell scenario without the input in one node), it follows that all such quantum strategies admit a left-local model in Eq.\eqref{fullmodel}. Hence, they cannot exhibit full network nonlocality.

We now provide details on the explicit measurement strategy found by numerically maximizing the violation of FNN with the measurement strategy in Eq.~\eqref{eq:FritzPOVMs}. 
We found out that the optimal measurements performed in each station of the network are projectors $\ket{\psi_{\theta}}\bra{\psi_{\theta}}$ restricted to qubits in the $x$-$z$ plane of the Bloch sphere $\ket{\psi_{\theta}}=\cos \theta \ket{0}+\sin \theta \ket{1}$ determined by the real number $\theta$. Additionally, this restriction is experimentally convenient, since such projectors can be readily implemented with standard qubit rotations about a single axis (e.g., they can be implemented in the photonic polarization degree of freedom by a single optical element like a half waveplate).

The measurement strategy at node $A$ can be fully characterized by two projection angles,
$\{\theta^A_{x=1},\; \theta^A_{x=2}\}$, which specify the two possible settings.
At node $B$, six projection angles are required, and are described by six parameters: $\{\theta^{B_0},\; \theta^{B_1}_{b_0=0}, \;\theta^{B_1}_{b_0=1}\}$ and $\{\theta^{B_0}_{b_1=0},\; \theta^{B_0}_{b_1=1},\; \theta^{B_1}\}$. These correspond to the single measurement defined in Eq.~\eqref{eq:FritzPOVMs}, which involves two triplets of projections on the subsystems $B_0$ and $B_1$. Each triplet encodes one possible feedback configuration: in the first, the measurement on $B_1$ depends on the outcome $b_0 \in \{0,1\}$ coming from the fixed measurement at $B_0$; in the second, the measurement on $B_0$ is conditioned on the outcome $b_1 \in \{0,1\}$ from the fixed measurement at $B_1$. Finally, at node $C$, we have two measurement settings for the projectors described by the two angles $\{\theta^C_{z=1},\; \theta^C_{z=2}\}$. 

In Table \ref{table:meas} we report the angles that best approximate (to the 4-th decimal digit) the optimal angles found numerically for the projectors that give the maximal violation value of FNN, via rational ratios of $ \pi$.

\begin{table}[ht!]
\begin{center}
\centering
\begin{tabular}{|l|l|l|l|l|}
\hline
 {\bf A} & {\bf B}0   & {\bf B1} &{\bf C } \rule[-1.3ex]{0pt}{1.3ex}\\
\hline
$\theta^A_{x=1}=-\frac{78}{183}\pi$ &$\theta^{B_0}= 0$ & $\theta^{B_1}= \frac{41}{103} \pi$& $\theta^{C}_{z=1}= \frac{\pi}{4}$  \rule[-1.3ex]{0pt}{1.3ex}\\
\hline
 $\theta^A_{x=2}= \frac{14}{17} \pi$& $\theta^{B_0}_{b_1=0}= \frac{19}{129} \pi $& $\theta^{B_1}_{b_0=0}= \frac{2}{27}  \pi$&$\theta^{C}_{z=2}= \frac{35}{108} \pi$ \rule[-1.3ex]{0pt}{1.3ex} \\
\hline
 & $\theta^{B_0}_{b_1=1}=\frac{9}{122} \pi$ &$\theta^{B_1}_{b_0=1}=\frac{23}{71}  \pi$& \rule[-1.3ex]{0pt}{1.3ex} \\
\hline
\end{tabular}
\end{center}
\caption{List of approximate angles as ratios of $\pi$, associated with the optimal single qubit projectors numerically found, showing FNN with our approach based on separable measurements.}
 \label{table:meas}
\end{table}

\section*{Appendix B: Details on MNN/FNN calculations}

In the following, we describe the optimization procedure we employed for the study of MNN and FNN of the feedback-based quantum strategy in Eq.\eqref{eq:MNNFritz}, defining a distribution $P(a,b,c|x,z)$.
In order to determine the belonging of $P(a,b,c|x,z)$ to the MNN or FNN set, we need to ensure its reproducibility by (semi-)local strategies with either one or both sources being classical.
We denote the possible sources configuration as right-, left-, or full-local, depending on whether the right, left, or both sources are constrained to be classical, respectively.
Similarly to~\cite{ciudad2024escaping}, to perform the optimization we use the fact that in the bilocality scenario any classical strategy can be described by an unpacked joint distribution $Q(a_1, \ldots, a_{|X|},b, c_1,\ldots, c_{|Y|})$, with the additional independence constraint $Q(a_0, \ldots, a_{|X|}, c_0,\ldots, c_{|Y|}) = Q(a_0, \ldots, a_{|X|}) Q(c_0,\ldots, c_{|Y|})$.
The original distribution can then be recovered by marginalization $Q(a_x,b,c_y) = P(a,b,c|x,z)$.

The optimization is performed by minimizing the amount of local noise $1-t$ needed to make $P(a,b,c|x,z)$ local, right-local, or left-local.
Specifically for the completely local case, we perform the following QCQP optimization:
\begin{align}
    \nonumber
    \max \;\; t &\\
    \nonumber
    \text{s.t.} \;\;& \exists \;\; \text{joint distribution}\;\;
        Q(a_0, \ldots, a_{|X|}, c_0,\ldots, c_{|Y|}) \\
        \nonumber
        &  Q(a_0, \ldots, a_{|X|}, c_0,\ldots, c_{|Y|}) = \\ &\quad= Q(a_0, \ldots, a_{|X|}) Q(c_0,\ldots, c_{|Y|})\,,\\
        \nonumber
        &Q(a_x,b,c_y) = t^2 P(a,b,c|x,z) +\\
        &\quad + \frac{1}{4}t(1-t) (P(a|x)+P(c|z))+\frac{1}{8}(1-t)^2\,.
        \label{eq:full-local}
\end{align}

Instead, for the left-local case, we solve this other optimization problem:
\begin{align}
    \nonumber
      \nonumber  \max \;\; t &\\ 
    \nonumber
    \text{s.t.} \;\; & \exists \;\; \text{joint distribution}\;\;  Q(a_0, \ldots, a_{|X|},b, c) \\
        \\
        \nonumber
        & Q(a_0, \ldots, a_{|X|}, c) =\\ &\quad = Q(a_0, \ldots, a_{|X|}) Q(c)\,,\\
        \nonumber
        &Q(a_x,b,c_y) = t^2 P(a,b,c|x,z)+ \\
        &\quad +\frac{1}{4}t(1-t) (P(a|x)+P(c|z))+\frac{1}{8}(1-t)^2\,,
        \label{eq:left-local}
\end{align}
where only the Alice part of the distribution has been ``unpacked''.
This effectively imposes the classical constraints only on the left source.
The optimization for the right-local case is done similarly.

Whenever $t=1$ we can conclude that the distribution is (semi-)local, while for lower values of $t$ the (semi-)local strategy is possible only when mixed with some amount of noise.
The parameter $t$ then effectively functions as a measure of robustness of (semi)-nonlocality.
The reason why $t$ has been chosen to be local for each source, leading to a quadratic constraint, was to make it a more realistic, albeit simplistic, noise model, instead of a mere robustness quantifier.

After solving the optimization problems numerically, using the Gurobi solver~\cite{gurobi}, for several parameters of our feedback-based strategy, we can categorize each distribution as MNN, if it is right- and left-local but not full-local, or FNN if it is none of the three.

\section*{Appendix C: Randomness certification in the bilocal network}
\label{app:random}
\begin{figure}[ht!]
    \centering
    \includegraphics[width=0.45\textwidth]{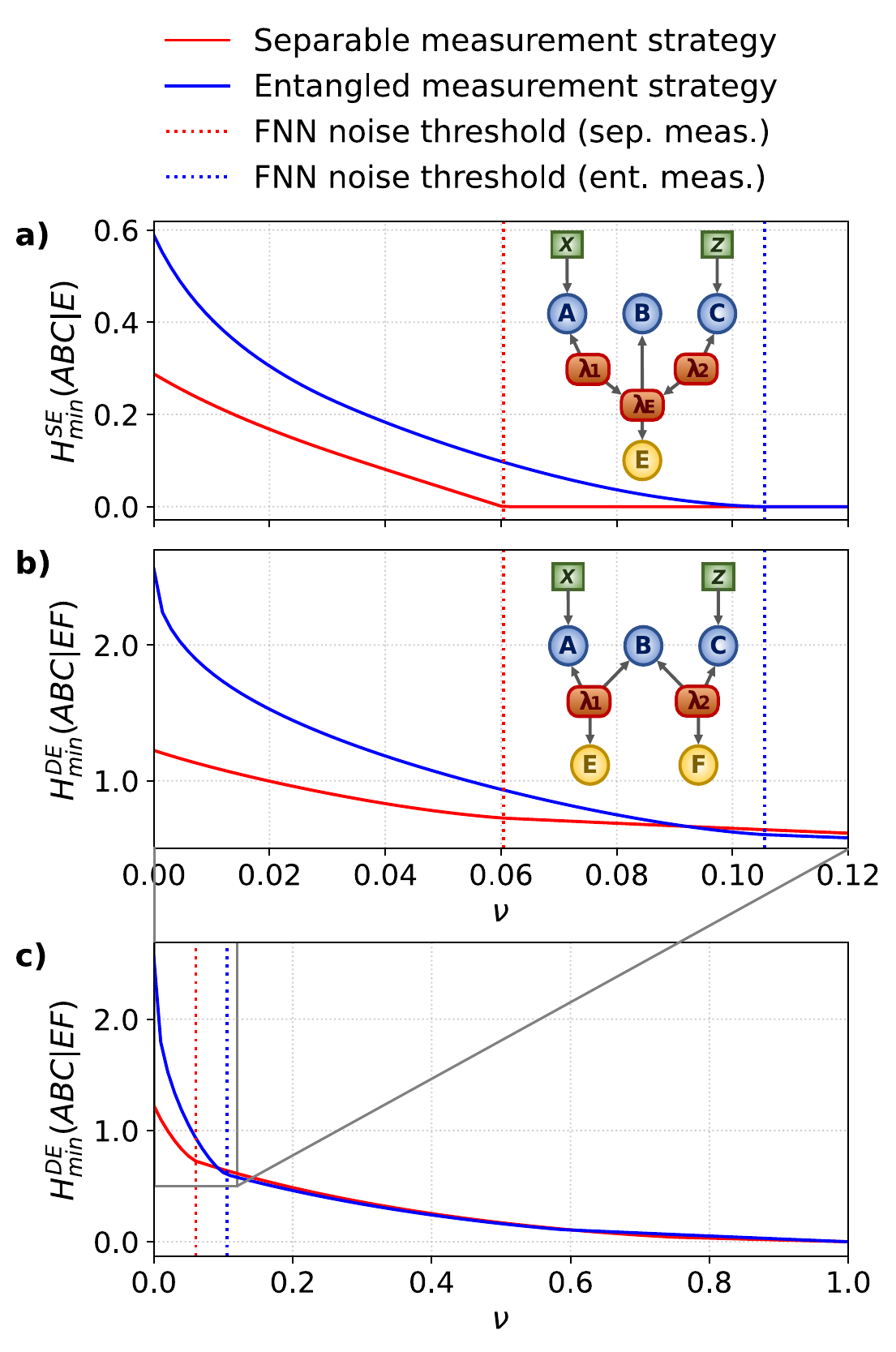}
     \caption{Randomness certification comparison between the entangled (blue curves) and entanglement-free (red curves) measurement schemes for FNN. Three-party min-entropy $H_{\mathrm{min}}^{\mathrm{SE/DE}}(ABC|E)$ as a function of the states' noise parameter $\nu$ in the \textbf{a)} SE and \textbf{b)} DE eavesdropper's configurations in the noise range with FNN. \textbf{c)} Min-entropy in the DE eavesdropper's configuration for the whole range of noise $\nu$.}
    \label{fig:rand_ABC}
\end{figure}

Device-independent randomness certification aims to guarantee the intrinsic unpredictability of measurement outcomes based solely on observed correlations, without relying on any assumptions regarding the internal functioning of the devices. The problem of bounding the eavesdropper's knowledge is defined as an optimization task over the set of valid quantum distributions that are compatible with the scenario under consideration. Specifically, we aim to maximize the guessing probability $P_{g}$, defined as the maximum probability that an eavesdropper can correctly predict the measurement outcomes.\\
This task has been extensively studied in the standard Bell scenario \cite{acin2016certified}, where a single source distributes correlations among two distant parties. In this case, thanks to the convexity of the correlation space, it is possible to cast the aforementioned optimization as a Semidefinite Programming (SDP) problem, exploiting the hierarchy of relaxation provided by the Navascués-Pironio-Acin (NPA) technique \cite{Navascues_2008}.\\
When multiple independent sources are present, as in network scenarios, causal independence relations may arise among the parties, making the set of correlations non-convex and, therefore, untreatable with the standard NPA technique. As shown in \cite{minati2026randomness}, the problem of randomness certification in networks can be tackled by using the so-called \textit{scalar extension} technique \cite{pozas2019bounding}, which allows us to take into account the independence relations inherent to the network structure.\\
In the context of the bilocal scenario, due to the presence of two independent sources, it is possible to consider a greater variety of eavesdropping strategies compared to the standard Bell-like scenarios. In particular, as described in the main text, we consider two different attacking strategies, respectively defined as the Strong-Eavesdropper (SE) and the Double-Eavesdropper (DE) scenarios. The SE scenario considers the most general attack within the bilocal network, in which a single eavesdropper has access to both the sources via an intermediate latent node. In this case, the probability that the eavesdropper correctly predicts the outcomes of the outer nodes' measurements is given by the following guessing probability:
\begin{equation}
    P_g^{\mathrm{SE}}(AC|E,x,z) = \sum_{a,b,c} p(a,b,c,e=(a,c)|x,z).
\end{equation}
On the other hand, in the DE scenario, two independent eavesdroppers, $E$ and $F$, are restricted to individual access to the sources. In this case, the guessing probability is defined as:
\begin{equation}
    P_g^{\mathrm{DE}}(A,C|E,F,x,z) = \sum_{a,b,c} p(a,b,c,e=a, f=c|x,z).
\end{equation}
Ultimately, the task of randomness certification in the bilocal network within a given eavesdropping scenario consists of solving the following optimization problem:
\begin{equation}
\label{eq:SDP_bilocal}
    \begin{aligned}
    \max_p \quad  & P_g^{\mathrm{SE/DE}}(AC|E(F),xz) \\
    \text{s.t.} \quad &p(a,b,c,e|x,z)  = \mathrm{Tr} (\rho_{ABCE} \cdot A_{a|x} \otimes B_{b} \otimes C_{c|z} \otimes E_{e}), \\
                      &p(a,b,c|x,z)   = \sum_e p(a,b,c,e|x,z) \\
                      &\rho_{ABCE}  = \rho_{AB_0E_0}\otimes\rho_{B_1CE_1}\;, 
    \end{aligned}
\end{equation}
in which the eavesdropper POVMs are constrained to have the form $E_e \otimes F_f$ when the DE scenario is considered. On the contrary, in the SE scenario, the action of the intermediate latent node $\lambda_E$ is accounted for by making both the POVMs $B_b$ and $E_e$ act on the Hilbert space $\mathcal{H}_{B_0} \otimes \mathcal{H}_{B_1} \otimes \mathcal{H}_{E_0} \otimes\mathcal{H}_{E_1}$. A non-zero amount of randomness is certified whenever this optimization problem returns a guessing probability lower than 1.\\
A more explicit figure of merit is given by the \textit{min-entropy}, defined as:
\begin{equation}
    H_{\mathrm{min}}^{\mathrm{SE/DE}}(AC|E(F), x,z) \equiv
- \log_2 \left( P_g^{\mathrm{SE/DE}} (AC|E(F),x,z)\right),
\end{equation}
which provides a lower bound on the amount of certified randomness directly in terms of random bits. In order to have a fair comparison in Fig.\ref{fig:rand}, for each configuration of SE/DE scenario and separable/entangled measurement in the central node, we considered the optimal pair of measurement settings $x$ and $z$, hence defining $H_{\mathrm{min}}^{\mathrm{SE/DE}}(AC|E) \equiv \max_{x,z} H_{\mathrm{min}}^{\mathrm{SE/DE}}(AC|E, x,z)$.\\
Following the same approach, we can also study how much randomness can be certified from all three observable nodes of the bilocal scenario. In this case, the quantities to be optimized are the following guessing probabilities:
\begin{align}
    &P_g^{\mathrm{SE}}(ABC|E,x,z) = \sum_{a,b,c} p(a,b,c,e=(abc)|x,z), \\
    &P_g^{\mathrm{DE}}(ABC|EF,x,z) = \sum_{a,b,c} p(a,b,c,e=(a,b), f=c|x,z).
\end{align}
The corresponding min-entropies, as a function of the noise parameter $\nu$, are reported in Fig.\ref{fig:rand_ABC}.
As we can see in Fig.\ref{fig:rand_ABC}a, in the SE scenario, the amount of certified randomness coincides with the one obtained from the outer nodes only. This is expected, since, within this attacking scenario, the eavesdropper can always have access to a copy of the state sent to $B$, thereby making $B$'s outcomes perfectly predictable.\\
This is no longer true in the DE scenario, in which the outcomes of $B$ contribute to the global amount of certified randomness,  as we can see from the larger min-entropies achieved in this case (reported in Fig.\ref{fig:rand_ABC}b) compared to the ones obtained from the outer nodes only (reported in Fig.\ref{fig:rand}b). However, this eavesdropper scenario has a peculiar behavior, since, as we notice in Fig.\ref{fig:rand_ABC}c, a non-zero amount of randomness is certified for any amount of noise. This is possible because the DE scenario requires $E$ to predict $B$'s outcomes, which depend on the state sent by source $\lambda_2$, i.e., a variable causally independent of $E$. Therefore, even if the noise does not allow for witnessing network nonlocality, there will be a "classical" contribution to the amount of certified randomness (further details are discussed in \cite{minati2026randomness}). Interestingly, in Fig.\ref{fig:rand_ABC}c, we can notice that the amount of certified randomness quickly increases for both the separable and entangled measurement strategies when the noise decreases below the corresponding thresholds. This indicates that, beyond the expected classical contribution, additional randomness is certified due to the nonlocal nature of the distribution.

%

\end{document}